\documentclass[aps,pra,twocolumn,floatfix]{revtex4}

\usepackage{amssymb}
\usepackage{amsbsy}
\usepackage{amsmath}
\usepackage[pdftex]{graphics}
\usepackage{graphics}
\usepackage{graphicx}
\usepackage{dcolumn}% Align table columns on decimal point
\usepackage{subfigure}
\usepackage{bm}% bold math
\def\be{\begin{equation}}
\def\ee{\end{equation}}

\begin{document}

\title{Transverse Ising Chain under Periodic Instantaneous Quenches: 
Dynamical Many-Body Freezing and Emergence of Slow Solitary Oscillations}

\author{Sirshendu Bhattacharyya} 
\affiliation{R.R.R. Mahavidyalaya,
Radhanagar, Hooghly, India}
\author{Arnab Das$^\ast$}  
\affiliation{
Theoretical Division (T-4), LANL, MS-B213,
Los Alamos, New Mexico - 87545, USA and \\
Max-Planck Institute for the Physics of Complex Systems, 
N\"{o}thnitzer Str. 38, Dresden 01187, Germany (present address).}
\author{Subinay Dasgupta}
\affiliation{
Department of Physics, University of Calcutta,
92 Acharya Prafulla Chandra Road, Kolkata 700009,
India}

\date{\today}

\begin{abstract}
We study the real-time dynamics of a quantum Ising chain driven periodically by 
instantaneous quenches of the transverse field (the transverse field varying 
as rectangular wave symmetric about zero). 
Two interesting phenomena are reported and analyzed:
{\bf (i)} We observe dynamical many-body freezing (DMF) \cite{AD-DQH}, 
i.e. strongly non-monotonic freezing 
of the response (transverse magnetization) with respect
to the driving parameters (pulse width and height) resulting
from equivocal freezing behavior of {\it all} the many-body modes.  
The freezing occurs due to coherent suppression of dynamics
of the many-body modes. For certain combination of the pulse height and period, maximal
freezing (freezing peaks) are observed. 
For those parameter values, 
a massive collapse of the entire Floquet spectrum occurs.  
{\bf (ii)} Secondly, we observe emergence of a distinct solitary 
oscillation with a single frequency, which can be much lower than the driving frequency. 
This slow oscillation, involving many high-energy modes, dominates the response remarkably in the limit of long observation time. 
We identify this slow oscillation as the unique survivor of destructive quantum interference
between the many-body modes. The oscillation is found to decay algebraically with time to a constant value.
All the key features are demonstrated analytically with 
numerical evaluations for specific results. 
 
\end{abstract}

\maketitle

\section{Introduction}
Dynamics of driven quantum many-body system is an emerging paradigm for 
studying and unveiling new quantum phenomena. 
Last few years have witnessed
a surge of theoretical endeavors in understanding dynamics of quantum many-body systems
under simple drivings. A major part of these recent activities is concentrated around 
quantum quenches, leading to several interesting and novel issues 
including (but not limited to) universal quench dynamics across quantum critical points
-- associated quantum Kibble-Zurek mechanism, physics of non-equilibrium excitations,
and the physics of thermalization in quantum systems (see for a review, Ref.\cite{Kris-RMP}; and C. de Grandi et. al., 
S. Mondal et. al., and U. Divkaran et. al, in Ref.\cite{ADBOOK} and 
references therein). 
The main focuses of these studies, e.g., the final defect density in a quantum quench, or
the effective temperature in a thermalized system, however, are insensitive to the 
details of the quantum coherence of the underlying many-body dynamics. For example, the dynamical idea 
behind quantum Kibble-Zurek mechanism \cite{KZ-Q}
is a robust translation of the classical Kibble-Zurek idea \cite{KZ-C} to quantum systems -- of course,
the origin of the relevant length scales and time scales are different.   

Here we focus on another important class of driven quantum non-equilibrium phenomenon,
where quantum coherence plays the central role. 
Though non-adiabaticity is a common covering for all interesting non-equilibrium
phenomenon, here {\it coherent} quantum mechanical suppression of dynamics
contributes crucially to the non-adiabaticity of the dynamics
which makes the resulting response behavior difficult to explain using classical
intuitions. We discuss dynamics of periodically driven quantum many-body system.
Coherent periodic driving
can give rise to surprising phenomenon in quantum many-body system, that counters our classical 
intuitions drastically \cite{AD-DQH}. The role of quantum coherence in 
the important context of superfluid-insulator transition   
realized in periodically driven optical lattice was demonstrated earlier \cite{Andre-1, Andre-1a}.  
Owing to the experimental break-through in
attaining long coherence time in quantum many-body systems in last decade, for example, within the
framework of atoms/ions in optical lattices and traps, this coherent regime is becoming
more and more accessible experimentally (see, e.g., \cite{Bogdan-Rev,Qsim,Kraus,Kim,Schatz}). 
Here we study the coherent dynamics (Schr\"{o}dinger dynamics at zero temperature) 
of a simple paradigmatic system - 
the transverse Ising chain \cite{BKC-Subir-SD} subjected to a train of rectangular pulses 
of the transverse field. Two interesting phenomenon are reported -- both purely
quantum mechanical in origin and are results of coherent many-body dynamics.

It has been observed recently that a class of integrable quantum many-body systems
exhibit the phenomenon of dynamical many-body freezing (DMF), i.e.  non-monotonic freezing
behavior of {\it all} the many-body modes when driven externally by varying a parameter 
in the Hamiltonian continuously \cite{AD-DQH}.
The said freezing behavior is  
counter-intuitive to the ``classical" picture of a driven system falling out of 
equilibrium. The classical behavior arises from competition of two timescales: 
the driving period and the relaxation time of the system (see, however, \cite{Tosatti}). 
In contrary to the expected monotonically increasing freezing behavior
of the system with respect to the driving frequency according to that picture,
we observe strongly non-monotonic freezing behavior, with
maximal freezing for certain combination of driving amplitude and frequency.
A related phenomenon, observed in context of a single particle 
localized in a periodically driven potential -- known as dynamical localization,  
or synonymously, coherent destruction of tunneling (CDT) 
is well studied \cite{Dunlap,Hanggi-1,Hanggi-2,Andre-2,Andre-3}. In \cite{Andre-3}, interestingly,
it has been shown in the context of periodically driven BEC, that the driving can 
lead to steady BEC-like states which are different from 
the equilibrium ground state of the undriven Hamiltonian.
Above findings motivate us investigating such phenomenon in a pulse driven many-body system, where,
instead of a smoothly varying driving rate, we have sequences of instantaneous
quenches and subsequent waiting times. Here we observe DMF, confirming the generality of the phenomenon
beyond sinusoidal driving. We deduce the exact condition for the maximal freezing analytically
and explore other characteristics of the freezing phenomenon.
 
In addition to DMF, we
observe another interesting phenomenon away from the freezing peaks.
In the limit of long observation time, we see spectacular dominance of a single long-lived
oscillation (with frequency much smaller than the driving frequency) 
in the response dynamics. Surprisingly, this happens even in the limit of strong and fast driving
(pulse amplitude and frequency much larger than the inter-spin coupling). 
We discuss the origin and nature of this intriguing quantum oscillation.

\section{The Model and the Dynamics}

We quench the transverse field $\Gamma$ from $+\Gamma_{0}$
to $-\Gamma_{0}$ and back in successive time intervals of duration $T$ 
in a transverse Ising  chain Hamiltonian:
\be \mathcal{H} = 
- J\sum_{j=1}^N s^x_j s^x_{j+1} - \Gamma(t) \sum_{j=1}^N s^z_j, 
\label{indepH}
\ee
\noindent
where the field $\Gamma(t)$ varies like a square-wave with period $T$  
at $t=0$ :
\be \Gamma(t) = \left\{ \begin{array}{rl}
       \Gamma_0 & \mbox{for $nT < t < (n+\frac12)T$}\\
     - \Gamma_0 & \mbox{for $(n+\frac12)T < t < (n+1)T$} \end{array} \right. \ee
with $n=0, 1, 2, \cdots$ and $\Gamma_{0}>0$. We set the energy scale by taking $J = 1$.    
In order to investigate the dynamics in this case, 
first we diagonalize Hamiltonian (\ref{indepH})
for a given value of $\Gamma$ 
by Jordan-Wigner transformation followed by Fourier transform
\cite{LSM}.
This transforms the Hamiltonian (\ref{indepH}) into a direct-sum of Hamiltonians of non-local 
free fermions of momenta $k$. The Hamiltonian preserves the parity of the 
fermion number (even/odd) and the ground state always lies in the even-fermionic sector. 
We work with the projection of
the Hamiltonian in this sector, given by  
\begin{eqnarray} 
\mathcal{H} &=& \bigoplus_{k>0}\mathcal{H}_k; \\ \nonumber 
 \mathcal{H}_k &=& (-2 i \sin k) \left[ a_k^{\dagger}a_{-k}^{\dagger} + a_k a_{-k} \right] \\ \nonumber
&-& 2(\Gamma + \cos k) \left[ a_k^{\dagger}a_k + a_{-k}^{\dagger}a_{-k} - 1 \right],   
\label{indepHk}
\end{eqnarray}
\noindent
where $k = (2n+1)\pi/N$;  $n = 0,1, ..., N/2-1$.  
The ground state of $\mathcal{H}_{k}$ is a linear combination
of the fermionic occupation number basis states $|0\rangle_{k} = |0_{k},0_{-k}\rangle$ 
(both $\pm k$ levels unoccupied) 
and $|1\rangle_{k}=|1_{k},1_{-k}\rangle$ (both $\pm k$ levels occupied), and the Hamiltonian does not
couple them with the two other basis states $|0_{k},1_{-k}\rangle$ and $|1_{k},0_{-k}\rangle$.
Hence starting with ground state, the dynamics always remains confined 
within a manifold which is the direct product of the $2-$dimensional  
subspaces spanned by $|0\rangle_{k}$ and $|1\rangle_{k}$.
We denote the eigenstates of $\mathcal{H}_k$ within this subspaces as 
$|(\Gamma, k)_-\rangle$ (ground state), 
and $|(\Gamma, k)_+\rangle$, with eigenvalues
$-\lambda(\Gamma, k)$, $\lambda(\Gamma, k)$, where 
\begin{eqnarray}
\lambda(\Gamma, k) & = & 2 \sqrt{\Gamma^2 + 1 + 2 \Gamma \cos k} \label{eigenvalue}\\
|(\Gamma, k)_-\rangle & = & i \cos \theta |1\rangle_{k} - \sin \theta |0\rangle_{k}  \label{eigenvector1} \\
|(\Gamma, k)_+\rangle & = & i \sin \theta |1\rangle_{k} + \cos \theta |0\rangle_{k} \label{eigenvector2} \\
\tan \theta & = & \frac{-\sin k}{\Gamma + \cos k + \sqrt{\Gamma^2 + 1 + 2 \Gamma \cos k} \label{theta_define}}.
\end{eqnarray}
\noindent
We now solve the Schr\"{o}dinger equation
\be  i\hbar \frac{\partial |\psi_k \rangle}{\partial t} = \mathcal{H}_k |\psi_k \rangle,  \ee
\noindent where the wave-function in the time-dependent energy eigen-basis may be expressed as 
\be |\psi_k \rangle = x_{-}(t) |(\Gamma, k)_{-}(t)\rangle + x_{+}(t)|(\Gamma, k)_{+}(t)\rangle.  \ee
If  $\Gamma(t) $ is constant (say, $\Gamma_0$) over a time interval $t_0$ to $t$, then we have
\be  x_{\pm} (t) =  x_{\pm} (t_0) \exp{\left\{ \mp \frac{i}{\hbar} (t - t_0) \lambda(\Gamma_0, k)\right\}}. \label{tdev}  \ee
\noindent
At time $t=0$ let the system be in a state
\be |\psi_k \rangle = \alpha |(\Gamma_0, k)_-\rangle  + \beta |(\Gamma_0, k)_+\rangle  \ee
with $|\alpha|^2 + |\beta|^2 = 1$. 
Then according to Eq. (\ref{tdev}) at 
$t = \frac{T}{2} - \epsilon$ (where $\epsilon$ 
is a small positive number), the coefficients are given by,
\be  \left( \begin {array}{l}
x_-(\frac{T}{2} - \epsilon) \\
x_+(\frac{T}{2} - \epsilon)  \end{array}      \right)= 
\left( \begin {array}{ll}
e^{i\mu_1} & 0 \\
0 & e^{-i\mu_1}  \end{array}      \right)
 \left( \begin {array}{l}
 \alpha \\
 \beta
 \end{array}  \right), 
\ee
where 
\be 
\mu_1 = \frac{T}{2\hbar} \lambda(\Gamma_0,k).
\label{mu-1}
\ee 
\noindent
Using the continuity of $|\psi_k \rangle$ at 
$t=\frac{T}{2}$ one obtains the wave function at $t=\frac{T}{2} +\epsilon$ in terms of
 $ |(-\Gamma_0, k)_- \rangle$ and $ |(-\Gamma_0, k)_+ \rangle$. 
Time evolution in the second half proceeds in the same way as in the first half and 
the transformation over one full cycle is given by,
\be  \left( \begin {array}{l}
x_-(T +  \epsilon) \\
x_+(T + \epsilon)  \end{array}      \right)=   {\bf U}_{k}
 \left( \begin {array}{l}
 \alpha \\
 \beta
 \end{array}      \right) \label {Ufirst} \ee
 where, 
\begin{eqnarray}  
{\bf U}_{k} = 
\left( \begin {array}{rr}
\cos \phi & -\sin \phi \\
 \sin \phi & \cos \phi  \end{array}  \right) 
\left( \begin {array}{ll}
e^{i\mu_2} & 0 \\
0 & e^{-i\mu_2}  \end{array}      \right) \\ \nonumber
\times \left( \begin {array}{rr}
\cos \phi & \sin \phi \\
- \sin \phi & \cos \phi  \end{array}      \right) 
\left( \begin {array}{ll}
e^{i\mu_1} & 0 \\
0 & e^{-i\mu_1}  \end{array}      \right).
\label{Udef}
\end{eqnarray}
\noindent Here
\be 
\mu_2 = \frac{T}{2\hbar} \lambda(-\Gamma_0,k) \quad {\rm and} \quad  
\phi = \theta_1 - \theta_2 
\label{mu-2}
\ee
where $\theta_1$, $\theta_2$ 
are the values of $\theta$ (as defined in Eq. (\ref{theta_define})) 
for $\Gamma = + \Gamma_0$, $-\Gamma_0$ respectively.

During the first half-cycle after $n$ full cycles, at a time 
$t = nT + \tau$ with $0 < \tau < \frac{T}{2}$, the coefficients are given by,
\be  \left( \begin {array}{l}
x_-(nT +  \tau) \\
x_+(nT + \tau)  \end{array}      \right)=  
\left( \begin {array}{ll}
e^{i\mu_3} & 0 \\
0 & e^{-i\mu_3}  \end{array}      \right)
  {\bf U}_{k}^n
 \left( \begin {array}{l}
 \alpha \\
 \beta
 \end{array}      \right) \label{Un1} \ee
 where $ \mu_3 = \frac{\tau}{\hbar} \lambda(\Gamma_0,k)$. 
Similarly, during the second half-cycle after $n$ 
full cycles, at a time $t = nT + \frac{T}{2} + \tau$, the coefficients are given by,
\begin{eqnarray}  
\left( \begin {array}{l}
x_-(nT +  \frac{T}{2} + \tau) \\
x_+(nT + \frac{T}{2} + \tau)  \end{array}      \right)=  
\left( \begin {array}{ll}
e^{i\mu_4} & 0 \\
0 & e^{-i\mu_4}  \end{array}      \right) \\ \nonumber
\times \left( \begin {array}{rr}
\cos \phi & \sin \phi \\
- \sin \phi & \cos \phi  \end{array}      \right)
\left( \begin {array}{ll}
e^{i\mu_1} & 0 \\
0 & e^{-i\mu_1}  \end{array}      \right)
  {\bf U}_{k}^n
 \left( \begin {array}{l}
 \alpha \\
 \beta
 \end{array}  \right) \label{Un2} 
\end{eqnarray}
 where $ \mu_4 = \frac{\tau}{\hbar} \lambda(-\Gamma_0,k)$. 
 
Transverse magnetization $M_z$ (per spin) at any time is given by,
\be M_z = -1 +  \frac{4}{N} \sum_{k=0}^{\pi} M_k = -1 + \frac{2}{\pi} \int_0^{\pi} M_k \, dk \label{Mz}\ee
where $M_k = \frac{1}{2}\langle \psi_k |  \left( a_k^{\ast}a_k + a_{-k}^{\ast}a_{-k} \right) |\psi_k \rangle$.
From Eqs. (\ref{eigenvector1}, \ref{eigenvector2}), 
\be M_k = | \left( x_- \cos \theta_j + x_+ \sin \theta_j \right) |^2 \label{Mk} \ee
with $j=1, 2$ according as we are in the first or second half-cycle respectively. 

In order to calculate $ {\bf U}_{k}^n$, giving the time-evolution after $n$th cycle,
we note that for any $2 \times 2$ matrix,
\[  {\bf U}_{k}^2 = -({\rm Tr}\,  {\bf U}_{k}) {\bf 1} + (\det {\bf U}_{k} )  {\bf U}_{k} \]
 This shows that one can write 
\be  {\bf U}_{k}^n = a_n {\bf 1} + b_n  {\bf U}_{k} \label{Unab} \ee
The recursion relations for $a_n$ and $b_n$ can be easily solved to 
get 
\begin{equation}
a_n = -b_{n-1} \quad \mathrm{and} \quad b_n = \sin(n\omega_k)/\sin \omega_k.
\label{an-bn}
\end{equation} 
\noindent
where $\cos \omega_k = \cos(\mu_1 + \mu_2) \cos^2 \phi + 
\cos(\mu_1 - \mu_2) \sin^2 \phi \label{cospsi}$.  The expressions $b_n$ are the 
Chebyshev polynomials of the second kind in $\cos \omega_k$.

\section{Dynamical Many-Body Freezing (DMF)}

%%%%%%%%%%%%%%%%%%%%%%%%%%
%\begin{figure}[htb]
\begin{figure*}
\includegraphics[width=0.8\linewidth,angle=0]{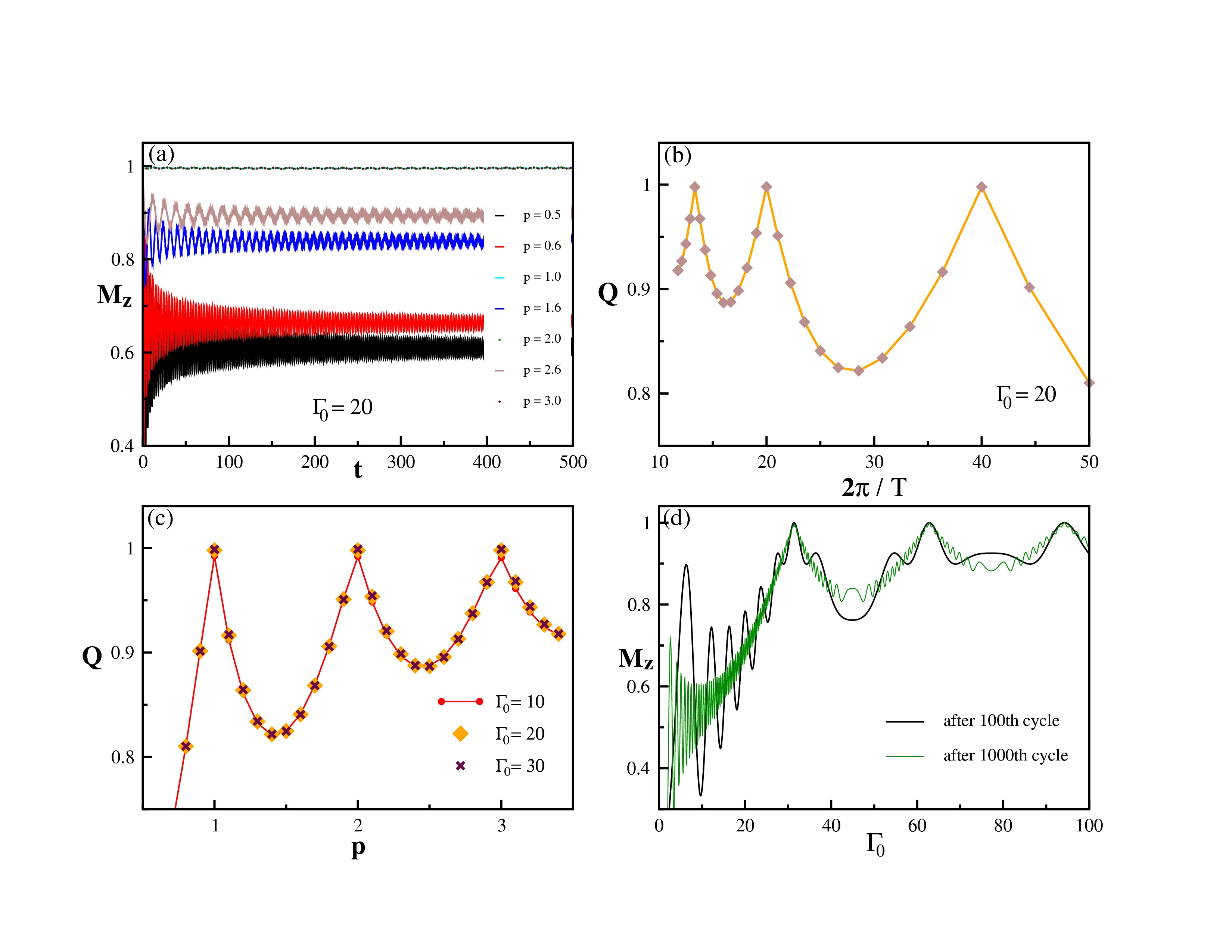}
\caption{ The dynamical many-body freezing (DMF) 
behavior of the resulting response.
{\bf (a)} Variation of $M_{z}$ with $t$ for different $p$($p=\frac{\Gamma_0 T}{\pi}$) for $\Gamma_{0} = 20$. 
{\bf (b)} Variation of $Q$ with $2\pi/T$ for $\Gamma_0 = 20$.
{\bf (c)} Variation of $Q$ with $p$ for different $\Gamma_0$. Maximal Freezing are seen for integer $p$. 
{\bf (d)} Magnetization after 100th and 1000th cycle at different $\Gamma_0$ for $T=0.1$. ($\hbar=1$)
}
\label{fig-1-new}
\end{figure*}
\noindent
The system is initially ($t=0$) in the ground state of the Hamiltonian with $\Gamma = +\Gamma_{0}$, before
it is driven by the pulses. We have computed the magnetization numerically at any time (within a cycle) by obtaining $x_-$ and 
$x_+$ from Eqs (\ref{Un1}, \ref{Un2}), substituting them in Eq. (\ref{Mk}) to get $M_k$ and then integrating it using Eq. (\ref{Mz}).
The result is presented in Fig. {\ref{fig-1-new}}.
Frame (a) and (b) shows that the response, 
i.e., the transverse magnetization $M_{z}$ which remains 
localized somewhere close to its initial value for all time.
In other words, the response retains the memory
of the breaking of the ${\mathcal Z}_{2}$ symmetry in transverse direction by the polarized
initial state through all later time, though the symmetry is respected by the driving over
each complete cycle.   
The degree of symmetry-breaking is given by the long-time average of $M_{z}$:
\begin{equation}
Q = \lim_{T_{f}\rightarrow\infty} \frac{1}{T_{f}}\int_{0}^{T_{f}} M_{z}(t) dt
\label{Q-factor} 
\end{equation} 
\noindent
$Q$ is also a measure of non-adiabatic freezing - if a driving were adiabatic, the
resulting response would always follow the field (i.e., trace the instantaneous ground state 
value of the response) and thus would preserve the symmetry of the Hamiltonian over a period.  
The maximum amplitude of oscillation of $M_{z}$ also determines the degree of freezing. 

\noindent
It is clear from Fig. {\ref{fig-1-new}} that for a given value of $\Gamma_{0}$, 
the non-adiabatic freezing $Q$ is a strongly non-monotonic function
of $T$.
When the condition 
\be \frac{\Gamma_{0}T}{\hbar} = \pi, 2\pi, 3\pi \cdots, \label{freeze}\ee
\noindent
is satisfied the freezing attains a maximum ($Q$ shows a peak), 
as shown in Fig. \ref{fig-1-new}(b) and \ref{fig-1-new}(c).
Naively speaking, for a given $\Gamma_{0}$, if $T$ is made larger, there is
more time for system to react to the successive flips made, and hence the
response is expected to be more adiabatic (smaller $Q$ and bigger response amplitude). This classical intuition
clearly does not hold in this case, as the freezing ($Q$ and response amplitude) is strongly non-monotonic in $T$
for a given $\Gamma_0$ (Fig.  {\ref{fig-1-new}} a). Strong maximal freezing 
of the entire many-body system ($Q$ peaks) observed for isolated points in the parameter 
space is also a surprising non-classical feature of DMF, arising from coherent quantum dynamics
\cite{AD-DQH}. 
  
In order to derive the extremal freezing condition (\ref{freeze}),
we set $\tau = 0$ and evaluate 
$M_k(t=nT)$ as a function of $n$. 
Thus, we are basically looking at the start of every oscillation. 
Also, we assume that initially (at $t=0$) the system was in the 
ground state for the transverse field at that moment. 
Thus, we set $\alpha=1$, $\beta=0$ in Eq. (\ref{Un1}), use Eq. (\ref{Unab}) 
there and obtain $x_-(nT)$ and $x_+(nT)$which is then substituted in Eq. (\ref{Mk}). The result is
\be M_k =  A_k + R_k \cos(2n\omega_k + \delta_k)  \label{MkAR} \ee
where 
\begin{eqnarray} 
A_k &=& \cos^2 \theta_1 + g_k f_k, \\ \nonumber
R_k^2 &=&  g_k^2  \left[ f_k^2 +  \sin^2(2\theta_1) \sin^2\mu_1 \sin^2 \omega_k \right], \\ \nonumber
\tan \delta_k &=& \frac{1}{f_k} \,\sin(2\theta_1) \sin\mu_1\sin \omega_k  \label{AkRk} 
\end{eqnarray}
with 
\begin{eqnarray} 
f_k &=& \sin(2\theta_1) \sin\mu_1 \cos \omega_k+  \sin(2\theta_2) \sin\mu_2   \\ \nonumber
g_k &=& \sin(2\phi) \sin(\mu_2)/(2\sin^2\omega_k) =|U_{12}|/(2\sin^2\omega_k)    \label{fkgk} 
\end{eqnarray}
and 
\be 
\omega_k =  \cos^{-1}\left[ \cos(\mu_1 + \mu_2) \cos^2 \phi 
+ \cos(\mu_1 - \mu_2) \sin^2 \phi \right ] \label{coswk} 
\ee
From Eq. (\ref{Mk}) and (\ref{MkAR}) we see,
the non-adiabatic freezing parameter $Q$ (Eq. \ref{Q-factor}) is given by
\begin{equation}
Q = -1 + \frac{2}{\pi}\int_{0}^{\pi} A_{k}dk
\label{Q-factor-2}
\end{equation} 
\noindent
Now, for large $\Gamma_0$, from Eqs. (\ref{mu-1}) and (\ref{mu-2})  we get
\begin{eqnarray} 
\phi  =  -\frac{\pi}{2}+\frac{\sin{k}}{\Gamma_0} 
+ \mathcal{O}\left (\frac{1}{\Gamma_0^3}\right), \;\;\; {\rm and} \\ \nonumber   
\mu_2 = \frac{\Gamma_0 T}{\hbar} \left[ 1 - \frac{\cos k}{\Gamma_0} 
+ {\mathcal O}\left(\frac{1}{\Gamma_0^2}\right) \right] 
\end{eqnarray}
The off-diagonal elements of the transfer matrix ${\bf U}_{k}$ becomes then,
\begin{eqnarray} U_{12} &=&  ie^{-i\mu_1}\sin \mu_2 \sin 2\phi \\ \nonumber 
&=&  - i.e^{-i\mu_1} \left[ \sin \left( \frac{\Gamma_0 T}{\hbar} \right)\frac{2\sin k}{\Gamma_0} 
+  {\mathcal O}(1/\Gamma_0^2) \right] = - U^{\ast}_{21} 
\label{U-Identity}
\end{eqnarray}
\noindent
Hence, according to Eq. (\ref{Udef}) if $\frac{\Gamma_0 T}{\hbar}$ 
is an integral multiple of $\pi$, ${\bf U}_{k}$ becomes a Identity matrix up to terms $1/\Gamma_0$
(since $\mu_{1} \approx -\Gamma_{0}T/\hbar$ for $\Gamma_{0} \gg 1$), 
and the system is found at the initial state (approximately) after each cycle. Note that the 
freezing occurs for {\em any initial state}, irrespective of whether it is an eigenstate
of the initial Hamiltonian or not. It is also consistent with the Floquet picture of quasi-energy degeneracy (see e.g.,
\cite{Mostafazadeh, Andre-Floquet}) employed in explaining dynamical localization. 
%applied to the $2\times2$ time evolution matrix ${\bf U}_{k}$ analyzed above. 
According to Floquet theory, the above time evolution operator ${\bf U}_{k}$, which induces 
evolution from $t=0$ to $t=T$, 
%should have the general form ${\bf U}_{k} = {\bf Z_{k}}(t)e^{i{\bf M_{k}}t}$,
should have the general form ${\bf U}_{k} = e^{i{\bf M_{k}}t}$,
%where ${\bf Z_{k}}(t)$ is $T-$periodic: ${\bf Z_{k}}(t+T)={\bf Z_{k}}(t)$ with ${\bf Z_{k}}(t=0)={\mathcal I}$ (Identity).
where ${\bf M_{k}}$ is a time independent Hermitian matrix (sharing same dimension and space as ${\bf U}_{k}$) 
\cite{Mostafazadeh}, with eigenvectors denoted by $|\mu_{1k}\rangle$ and 
$|\mu_{2k}\rangle$ corresponding to eigenvalues $\mu_{1k}$ and $\mu_{2k}$, which are the Floquet quasi-energies. 
Now as we have shown, in our case ${\bf U}_{k}$ tend to the Identity matrix up to term $1/\Gamma_{0}$ in large $\Gamma_{0}$
limit, which means both its eigenvalues $e^{i\mu_{1k}}, e^{i\mu_{1k}}$ tend to unity 
for every $k$ within the said approximation, resulting in a 
massive quasi-energetic degeneracy all over the many-body spectrum -- the crux of DMF.
Recently, another interesting manifestation of DMF is observed in periodically driven
bosons in optical lattice with small amplitude driving across the tip of the Mott lobe
\cite{Kris-DMF}.\\
   
The mechanism of DMF with rectangular driving can be visualized appealing to the simplicity
of the driving -- it consists of dynamics driven by piecewise time-independent Hamiltonians.
From Eq. (\ref{U-Identity}) one can see, 
the dynamics (in the eigen-basis $\{|(+\Gamma_{0},k)_{\pm}\rangle\}$ 
of ${\mathcal H}_{k}(+\Gamma_{0})$)
can be broken up into successive  
rotation of the basis by $\phi$ (due to successive flips of the transverse field) 
and intermediate accumulation of phases $\mu_{1,2}$
(due to intermediate waitings of length $T/2$). Clearly if one could adjust
the intermediate phases $\mu_{1,2}$ such that their effect is nullified {\it for all $k$},
in each cycle, then the system would return very closely to it's initial state after every cycle
for any of the eigenstates $\{|(+\Gamma_{0},k)_{\pm}\rangle\}$ as the initial state - the 
eigen-state of the initial Hamiltonian becomes Floquet states with degenerate quasi-energies
(albeit with a difference in sign, which does not matter in this case).
This happens, as explained in the paragraph following Eq. (\ref{U-Identity}), 
when $\Gamma_{0}$ is large and the condition for maximal freezing (Eq. \ref{freeze}) is satisfied.
For small $\Gamma_{0}$, $\mu_{1,2}$ would retain strong $k$-dependence, and hence this massive
collapse of Floquet spectrum would not have been possible (see Fig. \ref{fig-1-new} d).  
It is however worth noting that this simple picture of DMF cannot be extended in cases of 
continuously driven systems. For example, in the case of sinusoidal driving, the 
eigen-states of the initial Hamiltonian {\it do not} tend to return to themselves 
as one approaches the DMF freezing peak -- they retain a strongly $k$-dependent period 
($k$ is the quasi-momentum diagonalizing the initial Hamiltonian) 
of oscillation which actually diverges in the 
thermodynamic limit for certain modes as the DMF peak is approached. Freezing in that case
is visualized as vanishing of the {\it amplitude of oscillation} of each $k$ mode,
rather than ``all $k$-modes coming back to itself". It seems massive collapse of
the entire DMF spectrum is a result of integrability of the model and simplicity
of the driving.

\section{Long-lived Solitary Oscillation: The Survivor of Destructive Interferences}

\begin{figure*}
\includegraphics[width=0.8\linewidth,angle=0]{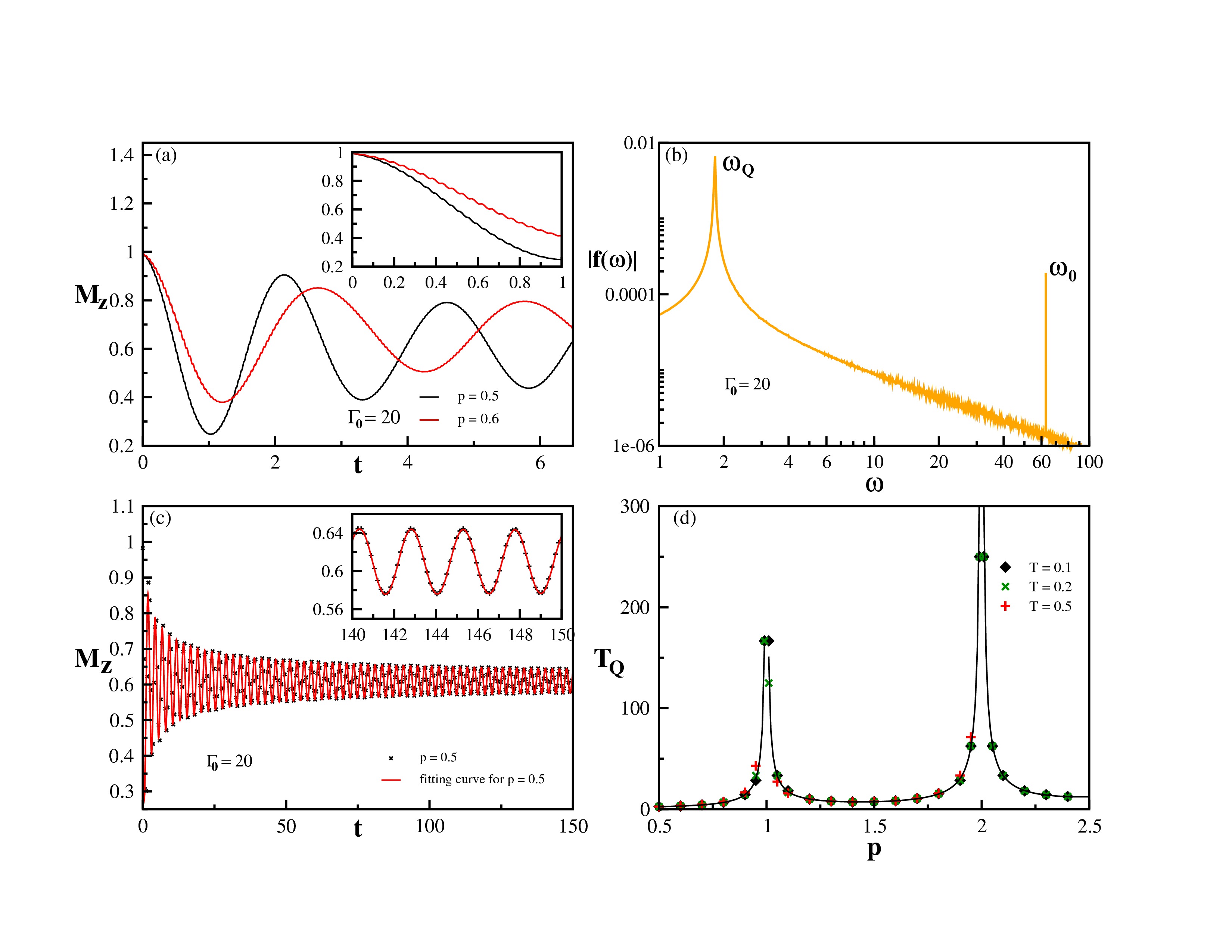}
\caption{Features of magnetization obtained by numerical integration of Eqs. (\ref{MkAR}) and (\ref{Mz}). 
{\bf (a)} Oscillations of two distinct time-scales are observed in $M_{z}(t)$. 
In addition to the expected one with period $T$ (matching to the driving), there is an additional longer
time-scale prominently visible.
{\bf (b)} Fourier transform of $M_{z}(t)$ showing two timescales visible in {\bf (a)}: 
two distinct peaks at angular frequencies $\omega_Q$ and $\omega_0$ are observed.
The peak  at $\omega_{Q} = 1.822$ in the Fig, matches quite 
well with the estimation (1.815) from Eq. (\ref{omegaQ}). 
While $\omega_{0} \approx 2\pi/T$, where $T=0.1$ is the driving period.  
{\bf (c)} Long-time behavior of $M_{z}(t)$ obtained numerically (points) 
and that given by Eq. fitting (\ref{Mzn_new}) (continuous curve) using is shown.
The envelop corresponding to the $1/\sqrt{n}$ decay (Eq. \ref{Mzn_new}) is visible.   
{\bf (d)} Variation of $T_Q = 2 \pi / \omega_Q$ with $p$ for $T=0.1$ 
obtained by Fourier transform. $T_Q$ tends to blow up (i.e., $\omega_{Q}$ vanishing up to order $1/\Gamma_{0}$) 
at integer values $p$ -- consistent with the observed maximal freezing at integer $p$. 
The points correspond to the values obtained from Fourier 
Transform and the continuous line is obtained from Eq. (\ref{omegaQ}). ($\hbar=1$)
}
\label{fig-2-new}
\end{figure*}
Analysis of $M_{z}(t)$ shows that it is dominated by a distinct solitary oscillation
in the long-time limit. The analysis of the response reveals 
sinusoidal oscillations of only two distinct  timescales --
one (denoted by $\omega_{0}$) matches with the driving period $T$ (as expected), while
the other, denoted by $T_{Q}$ (corresponding to frequency $\omega_{Q} = 2\pi/T_{Q}$), 
depends on all the driving parameters. $T_{Q}$ can be much larger compared to $T$. 
In spite of the fact that the driving has a large amplitude and high frequency,
and the system has several excitable energy levels, we observe
only one distinct non-trivial frequency in the response. 

This can be understood as follows. 
The transverse magnetization $M_{z}(t)$ (Eq. \ref{Mz}) at a time $t$ is a superposition of contributions 
$M_k$ for all $k$'s (Eq. \ref{MkAR}). 
For sufficiently large $n$, the argument 
$(2 n \omega_k + \delta_k)$ in Eq.(\ref{MkAR}) will be large (so that its cosine will fluctuate 
very rapidly with $k$) while $R_{k}$ will remain relatively slowly varying. 
Thus the contributions from neighboring $k$'s will cancel out due to destructive interference
(adding up with almost same amplitude but rapidly varying phase) over any small
intervals of $k$, except those around the stationary points of $\omega_{k}$ (with respect to $k$).
In the neighborhood of its stationary points,  $\omega_{k}$ is expected to vary slowly with $k$,
and hence the contribution from different $k'$s within such neighborhood is expected to add up
constructively. Elsewhere the contributions adds up destructively and can hence be ignored.
Thus we may write  
\be \int^{\pi}_0 R_k \cos (2 n \omega_k + \delta_k) dk \approx 
R_{\pi/2} \int^{\frac{\pi}{2} + \epsilon}_{\frac{\pi}{2} - \epsilon}  
\cos (2 n \omega_k + \delta_{\pi/2}) dk  \label{I1} \ee
By Taylor expansion of $\omega_k$  about the stationary point, we can write,
\be \cos (2 n \omega_k + \delta_{\pi/2}) =  \cos (2 n \omega_{\pi/2} 
+ \delta_{\pi/2}) \cos \left(nC\left[k - \frac{\pi}{2}\right]^2 \right) \label{Taylor_cos} \ee
where $C = (d^2\omega_k/dk^2)_{k=\pi/2}$. This finally gives,
\be M_z(n) \approx M_0 + \frac{a}{\sqrt{n}} \cos (n\omega_Q + \delta_{\pi/2}) \label{Mzn_new} \ee
where,
\be \omega_Q = 2 \omega_{\pi/2} 
=  2 \cos^{-1} \left\{ 1 - \cos^2 \phi [ 1 - \cos(\mu_1 + \mu_2)] \right\} \label{omegaQ} \ee
and
\[ M_0 = -1 + \frac{2}{\pi} \int_0^{\pi} A_k \, dk, \;\;\;\; a 
=  R_{\pi/2}\sqrt{\frac{\pi}{2C}}. \] 
%\int^{\infty}_{-\infty} \cos y^2 \, dy. \]
Above arguments are quite generic, and variants of them
can be found in different other contexts (see, e.g., \cite{Pomeau}).   
Survival of few such distinct oscillations of very long (compared to the driving period) time-scales
was also observed in an infinite-range transverse Ising model driven 
periodically in time \cite{Ad-Krish}. The results described in this section are  
manifestation of more general results regarding periodically driven quantum many-body system 
(\cite{AD-Floquet}). 

\noindent
We see from  Eq. (\ref{omegaQ}), when freezing condition ($p=$ integer) 
is satisfied, $\omega_Q$ vanishes for large $\Gamma_0$ 
up to terms linear in $1/\Gamma_0$ and $a \rightarrow 0$, $M_0 \rightarrow 1$. 
Numerical calculation of $M_z$ (using Eqs.(\ref{Un1}-\ref{Mk})) 
is presented in Fig.{\ref{fig-2-new}}. 
Discrete Fourier Transform of $M_{z}(t)$ also shows two peaks 
corresponding to $\omega_{0}$ and $\omega_{Q}$. The value of 
$\omega_Q$ obtained from there matches pretty well with 
the analytical expression in Eq.(\ref{omegaQ}).

Though our results are demonstrated for rectangular pulses, similar argument can be 
extended for other forms of periodic drivings. The only requirements for appearance of 
solitary oscillations (if they exist) are certain analytical properties of the 
response, continuity of the spectrum, and long driving time. Hence such oscillations 
are expected to appear quite generically in many-periodically driven coherent 
many-body quantum system, but analytical results might not be easy to extract in all cases. 
An extension of DMF for some other forms of periodic drivings 
may be achieved following \cite{Sacchetti}.

The phenomena we discussed above are result of quantum coherence. 
Further investigations in this direction are likely to reveal many new phenomena 
(see. e.g. Ref.\cite{Das-Moessner}). A natural open
question is whether they are realizable in real experiments, in presence of
the inevitable experimental imperfections existing within the present day setups.
Such experimental realizations would also allow for exploring these phenomena in
more generic non-integrable systems where accurate theoretical investigations 
could be difficult. Experimental observation of the above phenomena might be 
possible within the framework of coherent quantum simulation using trapped ions and atom in optical lattice.
In particular, DMF will have clear signature even for very small systems consisting of few spins
realized in the experimental systems above, since at the freezing peaks, {\it all} 
the momentum-modes freeze independent of system-size, 
whereas away from the peaks, considerable dynamics is expected for any system-size.
Coherent simulation of transverse Ising Hamiltonian with time-dependent transverse 
field, which can be varied adiabatically, has also been realized experimentally. 
In layered linear Paul traps using $^{171} Yb^+$ ions \cite{Kim}, and $^{25} Mg^+$ ions \cite{Schatz},
they realized transverse field Ising model where they can tune both spin-spin interactions
and transverse field with time.

\section{Summary}
We investigate dynamics of the transverse Ising chain under periodic
instantaneous quenches of the transverse field. 
We make two interesting observations -\\ 
\noindent
{\bf (i)} In the high amplitude ($\Gamma_{0} \gg J$) 
and fast quenching ($T \ll J $) limits we observe Dynamical
Many-body Freezing -- we see that the driven system freezes close to its
initial state, and the degree of freezing is a highly {\it non-monotonic}
function of the pulse amplitude $\Gamma_{0}$ and period $T$. The extremal
freezing is observed for $\Gamma_{0}T/\hbar = n\pi$ ($n=$ positive integers).
At these freezing ``peaks", the system remains frozen very strongly 
{\it independent} of its initial state. This freezing drastically contrasts
the classical notion of monotonic (with respect to the driving rate) 
freezing of a system under fast periodic driving -- a faster driving would give it
lesser time to react and hence would leave it more frozen. 
Quantum simulation of 
transverse Ising chain has already been realized experimentally -- the phenomenon should be
amenable to experimental verification within the said set-up and similar 
others for quantum simulation. \\
\noindent
{\bf (ii)} In the response dynamics, we observe emergence of a {\it single, distinct} 
timescale $T_{Q}$ (in addition to the time-scale of the driving) in the long-time limit.
This distinct oscillation decays much slower than other oscillations, following 
a $1/\sqrt{n}$ ($n=$number of sweeps) envelop. Dominance of a single
non-trivial frequency in the response is surprising, since the system 
is driven with pulses with high (compared to the intrinsic 
energy scale given by the spin-spin interaction $J$) amplitude and frequency.
We show that this surviving time-scale represents oscillations 
of the non-local momentum modes lying within a neighborhood of a unique 
point in the momentum space ($k = \pi/2$ here), where the contributions from the neighboring modes
adds up constructively. For all other parts of the momentum space such interferences are destructive, 
leading to mutual cancellation of oscillations of the neighboring modes.

\vspace{0.7cm}

%\noindent
%{\bf Acknowledgments:}

%\noindent
{\bf Acknowledgments:} 
The authors are grateful to Andre Eckardt, Joseph Samuel 
and Supurna Sinha for valuable comments.
AD acknowledges the support of U.S. 
Department of Energy through the LANL/LDRD Program. 
SD acknowledges financial support from CSIR (India).

\vspace{0.2cm}

\noindent
$^\ast$Correspondences should be addressed to AD (arnabdas@pks.mpg.de).

\end{document}